\documentclass[traditabstract,rnote]{aa}
\usepackage{graphicx}
\usepackage{txfonts}
\def\ltsim{\raise 2pt \hbox {$<$} \kern-0.6em \lower 4pt \hbox {$\sim$}}
\def\gtsim{\raise 2pt \hbox {$>$} \kern-1.1em \lower 4pt \hbox {$\sim$}}

\def\ltapprox{\raise 2pt \hbox {$<$} \kern-1.1em \lower 5pt \hbox {$\approx
$}}
\def\gtapprox{\raise 2pt \hbox {$>$} \kern-1.1em \lower 5pt \hbox {$\approx
$}}

\begin{document}

\title{
An unlikely radio halo in the low X-ray luminosity galaxy cluster RXC\,J1514.9-1523}

\author{S.~Giacintucci\inst{1},
D.~Dallacasa\inst{2,3},
T.~Venturi\inst{2},
G.~Brunetti\inst{2},
R.~Cassano\inst{2},
M.~Markevitch\inst{4},
R.~M.~Athreya\inst{5}
}
\institute
{Department of Astronomy, University of Maryland, College Park, MD,
  20742-2421, USA
\and
INAF- Istituto di Radioastronomia, via Gobetti 101, I-40129, Bologna,
Italy
\and
Department of Astronomy, University of Bologna, via Ranzani 1,
I-40127 Bologna, Italy
\and
Astrophysics Science Division, Laboratory for High
Energy Astrophysics, Code 662, NASA/Goddard Space Flight Center,
Greenbelt, MD 20771, USA
\and
Indian Institute of Science Education and Research,
  Central Tower, 
Sai Trinity Building, Sutarwadi Road, Pashan, Pune 411021, India
}
\date{Received 00 - 00 - 0000; accepted 00 - 00 - 0000}

\titlerunning{A giant radio halo in RXC\,J1514.9-1523}
\authorrunning{Giacintucci et al.}

\abstract{
We report the discovery of a giant radio halo in the galaxy cluster
RXC\,J1514.9-1523 at z=0.22 with a relatively low X-ray luminosity, 
$L_{X \, [0.1-2.4 \rm \, kev]} \sim 7 \times 10^{44}$ erg s$^{-1}$.
This faint, diffuse radio source is detected with the Giant
Metrewave Radio Telescope at 327 MHz. The source is barely detected 
at 1.4 GHz in a NVSS pointing that we have reanalyzed.
The integrated radio spectrum of the halo is quite steep, with a slope
$\alpha = 1.6$ between 327 MHz and 1.4 GHz. While giant radio halos
are common in more X-ray luminous cluster mergers, there is a less
than 10\% probability to detect a halo in systems with 
$L_X \ltsim  8 \times 10^{44}$ erg s$^{-1}$. The detection of a new
giant halo in this borderline luminosity regime can be particularly useful for discriminating
between the competing theories for the origin of ultrarelativistic
electrons in clusters. Furthermore, if our steep radio spectral index
is confirmed by future deeper radio observations, this cluster would
provide another example of the recently discovered population 
of ultra-steep spectrum radio halos, predicted by the model in which the
cluster cosmic ray electrons are produced by turbulent
reacceleration.}


\maketitle
%

\section{Introduction}

Hot, X-ray emitting gas is the dominant constituent of the intra-cluster 
medium (ICM) in galaxy clusters. The ICM is also permeated 
by magnetic fields and ultra-relativistic particles, whose energy densities 
and dynamical effects are still uncertain. The presence of these
non-thermal components is evidenced by giant, faint synchrotron radio halos, detected in 
the central $\sim$Mpc regions of a number of massive
clusters (e.g., Ferrari et al. 2008, Cassano 2009, Venturi 2011 for reviews).
Unlike radio galaxies, these diffuse, cluster-scale radio sources lack
any optical identification and are associated directly with the ICM,
in good spatial coincidence with the distribution of the hot, X-ray
emitting gas. They are produced by electrons with Lorentz 
factor $\gamma$ $\gtsim \, \, 1000$ spinning in large-scale $\mu$G 
magnetic fields. Their radio spectra are steep, with spectral indices
$\alpha>1$ (we adopt $S_{\nu} \propto \nu^{-\alpha}$, where 
$S_{\nu}$ is the flux density at the frequency $\nu$).

Observations show that radio halos are not common in galaxy clusters.
Large halos are found in only $\sim 1/3$ of the most massive and 
X-ray luminous clusters (e.g., Giovannini et al. 1999, Kempner \& Sarazin
2001, Venturi et al. 2008, Cassano et al. 2008), and become even rarer
in less massive systems. The rest of the clusters seems to form a distinct
population of {\em radio-quiet} systems (Brunetti et al. 2007, 2009). 

Clusters with and without a giant halo appear segregated 
in terms of their dynamical state: halos are located exclusively in 
merging systems, while clusters without radio halos are typically more
relaxed (e.g., Buote 2001, Cassano et al. 2010, and references
therein). Few exceptions are known, where a merging system does not
have a radio halo, 
typically with relatively low X-ray
luminosity ($L_X \ltsim 8 \times 10^{44}$ erg s$^{-1}$; e.g., 
Cassano et al. 2010, Russell et al. 2011).

The halo-merger connection suggests that the energy necessary to generate 
radio halos -- through acceleration of particles and amplification of 
magnetic fields -- is provided by cluster mergers.
Although the origin of radio halos is still debated (e.g., Brunetti et
al. 2008, Pfrommer et al. 2008, Donnert et al. 2010, Keshet and Loeb
2010, Brown and Rudnick 2011, Jeltema and Profumo 2011), 
current observations (e.g., Cassano 2009, and references
therein) appear to favour models where the giant halos 
are caused by merger-driven turbulence that reaccelerates 
relativistic particles ({\em reacceleration model}; Petrosian 2001, Brunetti
et al. 2001). In line with present data, these models predict that halos are 
more probably found in massive clusters and become quite rare in systems with mass 
$ \ltsim 10^{15}$ M$_{\odot}$ (i.e., $L_X \ltsim 7-8 \times 10^{44}$ erg
s$^{-1}$) at intermediate redshift ($z \sim 0.2 \div 0.5$; Cassano et al. 2008).

Here, we report the discovery of a giant radio halo in
RXC\,J1514.9-1523, a galaxy cluster at z=0.22 with a 
relatively low X-ray luminousity,  $L_{X \,[0.1-2.4 \rm \, kev]}=7.2 \times 10^{44}$ 
erg s$^{-1}$ (B\"{o}hringer et al. 2004).

We adopt the $\Lambda$CDM cosmology with H$_0$=70 km s$^{-1}$ Mpc$^{-1}$, 
$\Omega_m=0.3$ and $\Omega_{\Lambda}=0.7$. At the redshift of 
RXC\, J1514.9-1523, this gives a scale of 
$1^{\prime \prime}=3.59$ kpc.

\section{Radio observations}

RXC\,J1514.9-1523 is in the cluster sample of the extended {\em Giant
  Metrewave Radio Telescope} ({\em GMRT}) Radio 
Halo Survey, an ongoing project to expand the existing sample of
galaxy clusters (Venturi et al. 2007 \& 2008; see Cassano et 
al. 2010 for details). In this work, we analyze
archival {\em GMRT} observations of RXC\,J1514.9-1523 at 327 MHz and
{\em Very Large Array} ({\em
VLA}) data at 1.4 GHz from the NVSS\footnote{NRAO {\em VLA} Sky Survey (Condon et al. 1998).} pointing
containing the cluster, that we have reprocessed. We also 
observed the cluster with the {\em GMRT} at 235 MHz/610 MHz
in August 2009, as part of the extended {\em GMRT} Radio 
Halo Survey. 
Variable radio frequency interference was present for
the whole observation, particularly on the shortest spacings. The
subsequent editing left a sparse sampling 
at the short baselines, preventing the production of an image 
of the radio halo of any use.
A reobservation is currently planned.

\subsection{{\em GMRT} observations at 327 MHz}

RXC\,J1514.9-1523 was observed with the {\em GMRT} at 327 MHz 
in April 2004 as part of the {\it GMRT} Cluster Key Project (05VKK01), 
for a total of $\sim$140 minutes on source. The data were recorded 
using both the upper and lower side bands (USB and LSB), providing a 
total observing bandwidth of 32 MHz. The default spectral-line mode 
was used. The USB and LSB datasets were calibrated individually using
the NRAO Astronomical Image Processing System
(AIPS) package, following the procedure described in Giacintucci et
al. (2008). After bandpass calibration and a priori amplitude
calibration, a number of phase-only self-calibration cycles and 
imaging were carried out for each data set.  We initially
self-calibrated the longest baselines using only the point
sources in the field, then we progressively included short baselines and
resolved radio sources. Finally, we included also the emission from the
radio halo. Wide-field imaging was implemented in each step of the
data reduction. We used 25 facets to cover a total field of $\sim$2.7$^{\circ}\times2.7^{\circ}$.
The USB and LSB data sets were then combined together 
to produce the final images using the multi-scale CLEAN 
implemented in IMAGR, which allows better imaging of extended sources
compared to the traditional CLEAN (e.g., Clarke \& Ensslin 2006; for a
detailed discussion see Appendix A in Greisen, Spekkens \& van Moorsel
2009). We used delta functions as model components for the
  unresolved features and circular Gaussians for the resolved ones,
with increasing width to progressively highlight the extended emission during the clean.
The rms sensitivity levels achieved in the image at full resolution
($\sim$14$^{\prime \prime}$) is $\sim$120 $\mu$Jy beam$^{-1}$.
We also produced images with lower resolutions,
down to $\sim$50$^{\prime \prime}$,
by tapering the $uv$ data using UVTAPER and ROBUST in IMAGR. 
The noise reached in the images at the lowest resolution is $\sim$0.3 
mJy beam$^{-1}$.

The flux density scale was set using the scale of Baars et al. (1977).
The average residual amplitude errors are $\ltsim$8\% (e.g., Chandra et al. 2004).

\begin{figure*}
\includegraphics[width=8.8cm]{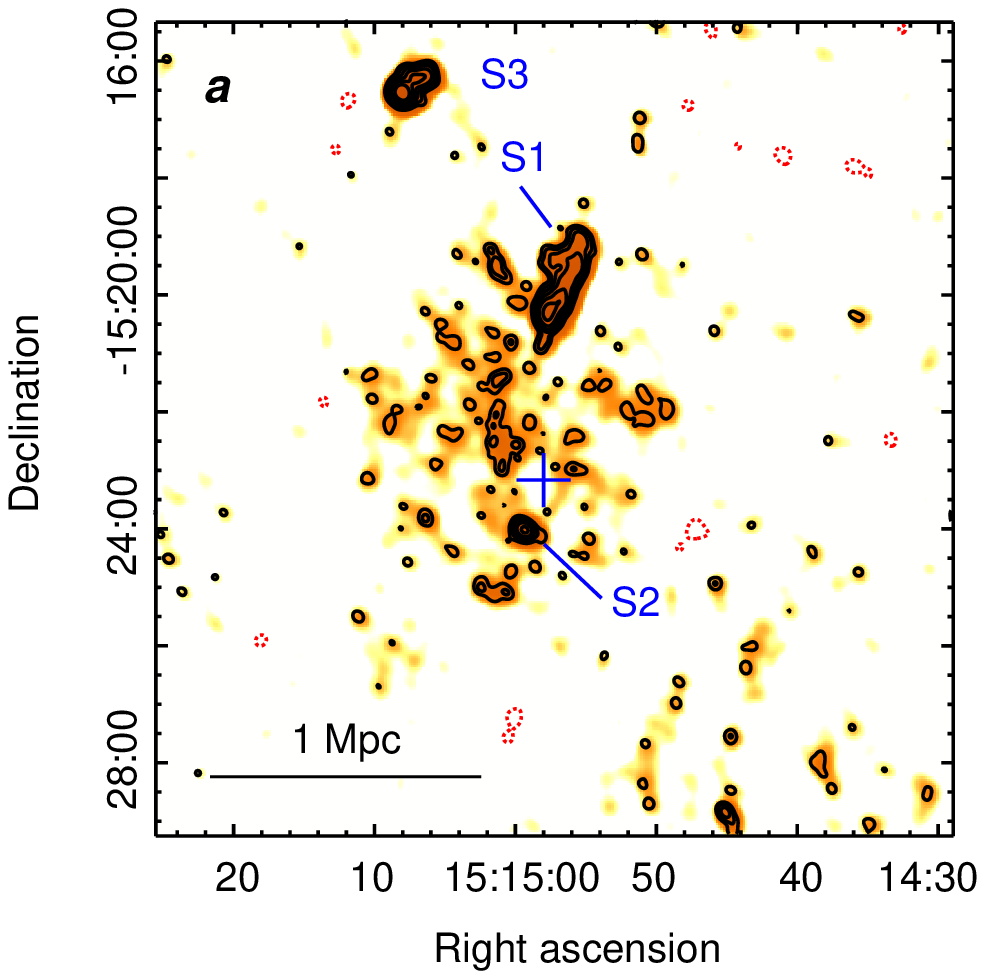}
\includegraphics[width=8.8cm]{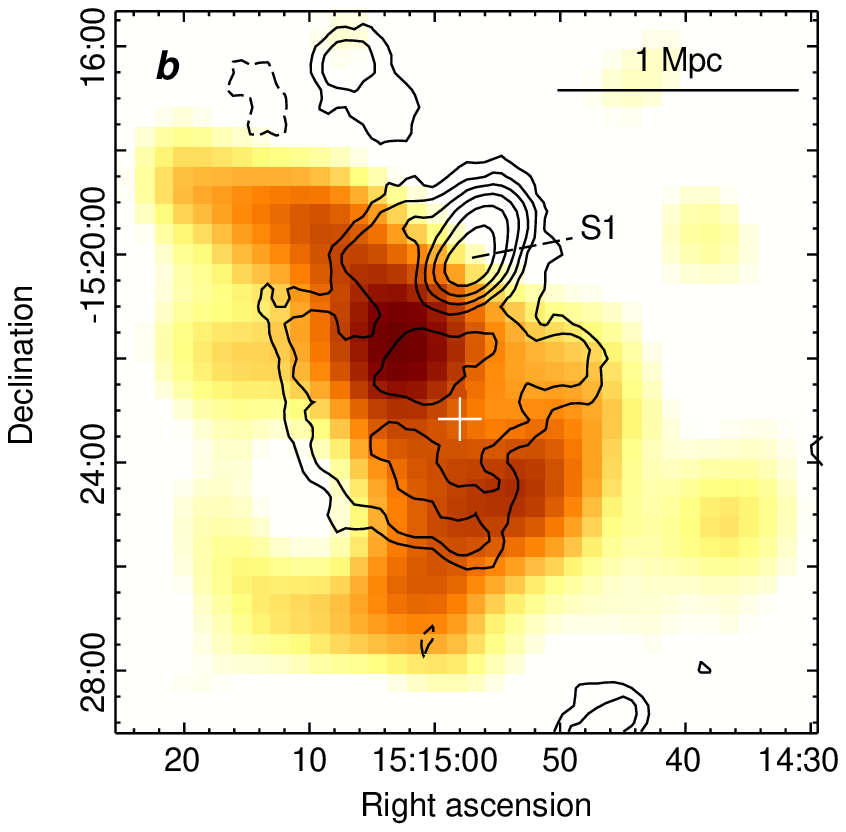}
\caption[]{{\it a:} {\em  GMRT} full-resolution image at 327 MHz (contours and colors).
The restoring beam is $14^{\prime\prime}\times 12^{\prime\prime}$. Black
contours start at +3$\sigma$=0.36 mJy beam$^{-1}$ and 
then scale by a factor of 2. The $-3\sigma$ level is shown as red,
dashed
contours. {\it b:} {\em GMRT} 327 MHz contours, after subtraction of 
S2 and S3 (panel {\em a}), overlaid on the smoothed RASS X-ray image. 
The restoring beam is $53^{\prime\prime}\times 41^{\prime\prime}$ and 
the rms noise level is 0.27 mJy beam$^{-1}$.
Contours start at +3$\sigma$ and then scale by a factor of 2. 
The $-3\sigma$ level is shown as black dashed contours. 
The cross shows the position of the cluster X-ray centre. }
\label{fig:327}
\end{figure*}

\begin{figure*}
\centering
\includegraphics[width=18.5cm]{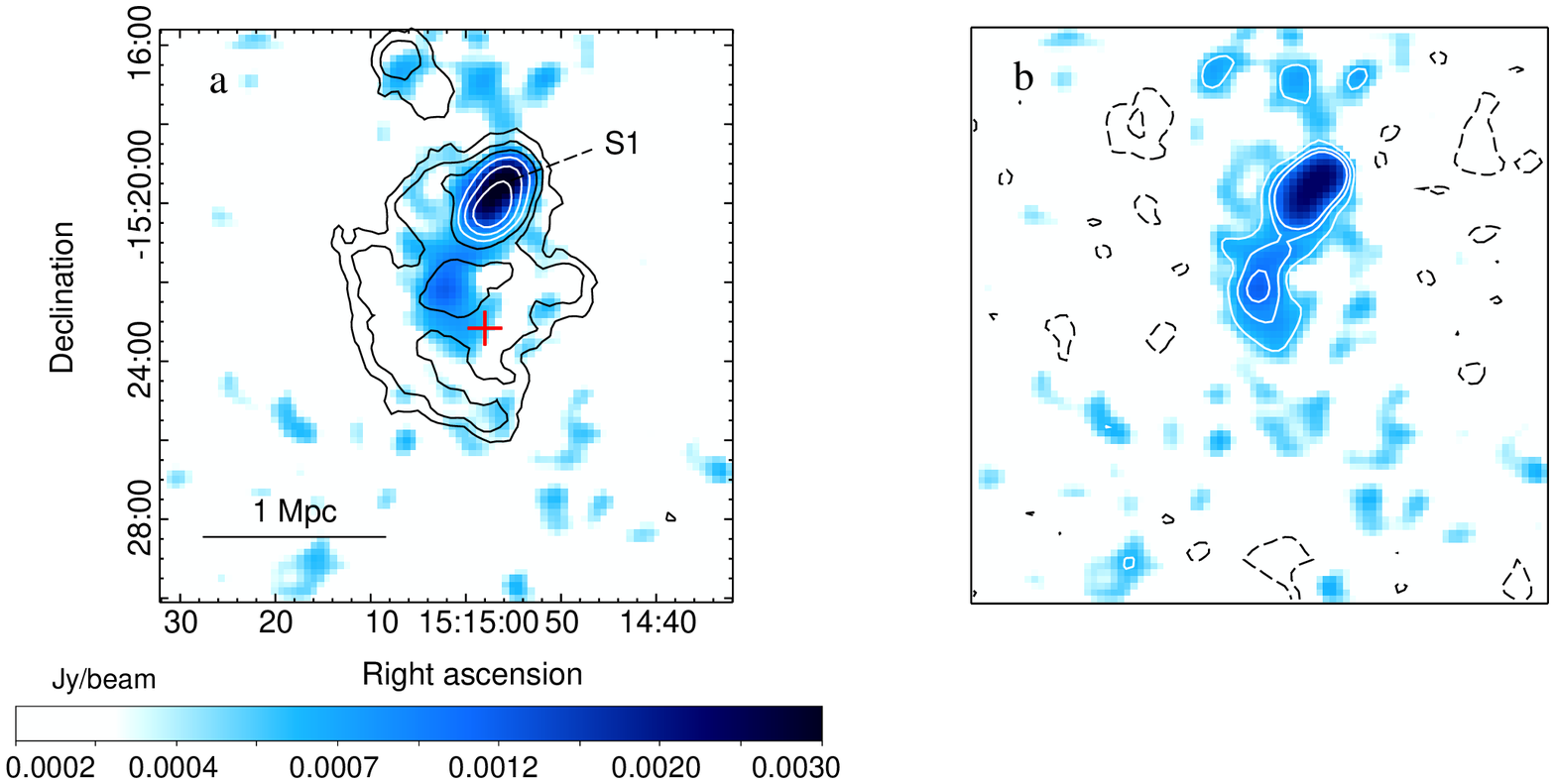}
\caption[]{{\it a:} {\em GMRT} 327 MHz contours of the radio halo (black and
  white; same as Figure 1{\em b}),
overlaid on the {\em VLA} 1.4 GHz image at the resolution of
$42^{\prime\prime}\times 37^{\prime\prime}$, obtained from 
the re-processed NVSS data. The rms noise in the 1.4 GHz
image is 0.25 mJy beam$^{-1}$. Point sources,
derived from an image produced with only baselines 
$\ge 2$ k$\lambda$, have been subtracted out at 1.4 GHz.
{\it b:} Same re-processed NVSS image as panel {\it a}), with 1.4 GHz 
contours overlaid at levels -0.5 (dashed black),
0.5, 0.75 and 1 (white) mJy beam$^{-1}$. 
}
\label{fig:nvss}
\end{figure*}

Figure 1a presents the 327 MHz image of the $\sim12\times12$ 
arcmin$^{2}$ field containing RXC\,J1514.9--1523 at full resolution ($\sim$14$^{\prime \prime}$).
The field corresponds to an area of $\sim$2.6$\times$2.6 Mpc$^{2}$ at
the redshift of the cluster.
The cross shows the position of the cluster X-ray centre, as reported
in the REFLEX\footnote{{\em ROSAT}-ESO Flux Limited X-ray (B\"{o}hringer et al. 2004) } cluster 
catalogue (RA=15h14m58.0s, DEC=$-$15$^{\circ}$23$^{\prime}$10$^{\prime
  \prime}$). 
The brightest sources in the cluster
region, S1 and S3, are both extended with tailed morphology. S2,
south of the cluster centre, is unresolved.
Faint, diffuse emission is observed in the region between the
head-tail S1 and S2. Figure 1b shows the
327 MHz image restored with a beam of $53^{\prime \prime} \times
41^{\prime \prime}$, after subtraction of the clean components associated with all
individual radio sources in the cluster area but S1.
The radio contours are superposed on the smoothed 
{\em ROSAT} All-Sky-Survey (RASS) X-ray image. 
The image confirms the existence of large radio halo at the 
cluster centre, with a size of $\sim$1.4 
Mpc and a fairly roundish morphology. The halo
covers a large part of the complex X-ray emission 
associated with the cluster, which is mostly elongated in 
the northeast-southwest direction, with two possible
peaks, probably due to an ongoing cluster merger. The brightest
portion of the halo is approximately coincident with the main 
X-ray peak.

In Figure 2a, we overlay the 327 MHz contours of the radio halo
on the NVSS {\em VLA} image at 1.4 GHz at the resolution of
$42^{\prime\prime}\times37^{\prime\prime}$, also shown as 
contours in Fig.~2b. The 1.4 GHz image has been obtained 
after new calibration of the NVSS pointing containing
RXC\,J1514.9-1523. While only hints of the radio halo were 
barely visible on the public NVSS image, the new data calibration, 
joint with phase self-calibration, allowed us to obtain an
image of the diffuse source. We first used only baselines longer than 2 k$\lambda$ to image 
the discrete radio sources. The clean components associated 
with these sources were then subtracted from the $uv$-data, and
the resulting data set was used to obtain the image in Figure 2.

The radio halo is quite faint at 1.4 GHz; only the central, bright
region of the emission seen at 327 MHz is detected in the 
reprocessed NVSS image, at a level starting from $\sim$0.5
mJy beam$^{-1}$ with a peak of $\sim$1.2 mJy beam$^{-1}$.

The flux density of the halo at 327 MHz is 102$\pm$9 mJy.
This was obtained by integration of the low-resolution image
in Fig.~1b over the area that encompasses the whole diffuse
emission, and after subtraction of the contribution of S1,
measured at full resolution (81$\pm$6 mJy).
The flux density at 1.4 GHz is $10\pm2$ mJy (after subtraction 
of 38$\pm$2 mJy associated with S1). The resulting spectral index of the halo 
is $\alpha = 1.6\pm0.2$. This value is 
the average over the whole halo extent at 327 MHz, but 
the size of the emission detected in the NVSS pointing 
is considerably smaller and part of the halo flux density 
may be missed. Deeper and higher resolution observations 
at 1.4 GHz are therefore needed to accurately measure 
the total flux density of the halo and place a better constraint on
its spectral index.

\section{Discussion}

The 327 MHz images presented in Section 3 show that 
RXC\,J1514.9-1523 possesses a large, central radio halo. 
The diffuse source is brighter and larger at 327 MHz 
than in the reprocessed NVSS image at 1.4 GHz, where the halo is 
only marginally visible. 

With an X-ray luminosity of $L_{X \, \rm [0.1-2.4 \, kev]}=7.2 \times
10^{44}$ erg s$^{-1}$, RXC\,J1514.9-1523 is one of the faintest
systems among the radio halo clusters at redshift larger than 0.2 
(e.g., Brunetti et al. 2009). Based on the {\em GMRT} Radio Halo Survey 
at 610 MHz, coupled with literature
information, Cassano et al. (2008) established that the fraction of
radio halo clusters in highly massive systems (i.e., with $ L_X > 8 \times
10^{44}$ erg s$^{-1}$) is $\sim 1/3$, while
the occurrance of halos in less X-ray luminous clusters
is less than $10\%$ at intermediate redshifts ($z\sim 0.2\div 0.5$). 
Therefore, the detection of a giant halo in a relatively low X-ray
luminousity cluster at $z>0.2$, such as RXC\,J1514.9-1523, 
adds important information to the statistics of halos in such systems.
 
The decrease of the fraction of radio-halo clusters 
with decreasing mass and X-ray luminosity of the cluster is a
prediction of turbulent reacceleration models,
which expect mergers between massive clusters 
( $\gtsim$ 1-2 $\times 10^{15}$ M$_{\odot}$) 
to be more efficient accelerators of relativistic particles
(Cassano \& Brunetti 2005, Cassano et al. 2006).
In these models, the synchrotron spectra of radio halos should 
have a spectral break at a frequency which depends on the fraction 
of turbulence energy converted into electron reacceleration. 
Only collisions between massive clusters with similar-mass subclusters 
are able to generate radio halos with a break frequency $\ge$1 GHz. 
The formation of this type of halos in clusters with lower masses, or
during moderately energetic mergers, is more difficult and rare. 
Here, less efficient reacceleration may result in halos with a 
spectral break at frequencies lower than 1 GHz. Such halos would be
barely detectable at $\sim$ 1 GHz, but bright at lower frequencies,
and should have an extremely steep spectrum when observed
at GHz frequencies. 

Interestingly, most of the unrelaxed clusters without detected radio
halo have $ L_X \ltsim 8 \times 10^{44}$ erg s$^{-1}$ (e.g.,Cassano et al. 
2010a, Russell et al. 2011). These {\em outliers} may be detected 
in the radio band by future observations at lower frequencies, 
for instance with the {\em LOw Frequency ARray} ({\em LOFAR}) 
and {\em Long Wavelength Array} ({\em LWA}). In one of these systems, 
A\,781, large-scale residual emission at 325 MHz has been recently
reported by Venturi et al. (2011), which might indeed trace an
underlying halo with very steep spectrum (but see Govoni et al. 2011).

The detection of halos with very steep spectra ($\alpha > 1.5$) 
disfavours a secondary origin of the radio emitting electrons
for energy reasons
(Brunetti 2004, Pfrommer \& Ennslin 2004). 
In hadronic models, clusters hosting halos with very steep spectrum
contain an untenable large population of cosmic ray protons (e.g.,
Brunetti et al. 2008, Macario et al. 2010), 
while a much smaller energy is constrained by present upper limits in
the gamma-ray band (Aharonian et al. 2009, Ackermann et al.
2010, Jeltema \& Profumo 2011).
If the steep radio spectral index ($\alpha=1.6$) of the halo 
is confirmed by future deeper observations, RXC\,J1514.9-1523 would 
provide another example of an
ultra-steep 
spectrum radio halo. 
Interestingly, RXC\,J1514.9-1523 shares similar properties with the
prototype of this population of halos, A\,521 (Brunetti et al. 2008, Dallacasa et al. 2009).
The X-ray luminosity of the two clusters is 
almost identical, and so are the radio powers of the halos with log$P_{\rm 1.4 \, GHz} \rm
(W \, Hz^{-1}) = 24.23$ and 24.18, respectively. Indeed, the two
clusters occupy the same region in the radio-X-ray luminosity 
diagram for giant radio halos (e.g., Brunetti et al. 2009). In
particular, as A\,521, RXC\,J1514.9-1523 appears slightly underluminous 
in the radio band with respect to the $P_{\rm 1.4 \, GHz}-L_X$ correlation for halos,
possibly due to its steep spectrum.

\section{Conclusions}

Using {\it GMRT} data at 327 MHz, we found a Mpc-size radio halo
in the galaxy cluster RXC\,J1514.9-1523. With an X-ray luminosity 
of $\sim 7 \times10^{44}$ erg s$^{-1}$, this cluster is one of the 
less X-ray luminous systems with a radio halo at $z > 0.2$. 
Observations show that the percentage of clusters with 
a halo decreases with decreasing cluster mass. In particular, 
giant halos are found only in less than 10\% of clusters with 
$L_X \ltsim 8 \times10^{44}$ erg s$^{-1}$. A drop of the fraction of
clusters with giant halos is expected at these X-ray luminosities
in the context of turbulent reacceleration models.
The discovery of a giant halo in a cluster in this borderline
X-ray luminosity regime adds therefore a useful piece to the 
puzzle of the origin of radio halos, as it adds information to the 
statistics of halos in an important range of cluster masses.
In addition, if the steep spectrum of the halo is confirmed by 
future radio observations, RXC\,J1514.9-1523 will also 
provide crucial support to reacceleration models and 
contribute to answer the long standing question of the origin 
of cluster-wide radio halos.
\\
\\
{\it Acknowledgements.}  
We thank the staff of the {\em GMRT} for their help during the observations.
{\em GMRT} is run by the National Centre for Radio Astrophysics of the Tata
Institute of Fundamental Research. SG
acknowledges the support of NASA through Einstein Postdoctoral
Fellowship PF0-110071 awarded by the {\em Chandra}
X-ray Center (CXC), which is operated by SAO. The National Radio
Astronomy Observatory is a facility of the National Science 
Foundation operated under cooperative agreement by Associated Universities, Inc.

\end{document}